\documentclass[letterpaper,12pt]{article}

\usepackage{osajnl2}
\usepackage{amsmath}
\usepackage{subfig}
\usepackage{grffile}
\usepackage{float}
\usepackage{graphicx}

\graphicspath{%
  {../figs/illustrations/}%
  {../figs/input/}%
  {../figs/results/}%
}

\begin{document}

\title{Designing and using prior knowledge for phase retrieval}

\author{Eliyahu
  Osherovich$^{*}$, Michael Zibulevsky, and Irad Yavneh}
\address{Computer Science Department, Technion ---
  Israel Institute of 
  Technology, \\ Haifa, 32000, Israel} \address{$^*$Corresponding
  author: oeli@cs.technion.ac.il}

\maketitle{}

\begin{abstract}
  In this work we develop an algorithm for signal reconstruction from
  the magnitude of its Fourier transform in a situation where some
  (non-zero) parts of the sought signal are known. Although our method
  does not assume that the known part comprises the boundary of the
  sought signal, this is often the case in microscopy: a specimen is
  placed inside a known mask, which can be thought of as a known light
  source that surrounds the unknown signal. Therefore, in the past,
  several algorithms were suggested that solve the phase retrieval
  problem assuming known boundary values~\cite{hayes82importance,hayes83recursive,fienup83reconstruction,fienup86phase}.
  Unlike our method, these methods do rely on the fact that the known
  part is on the boundary.

  Besides the reconstruction method we give an explanation of the
  phenomena observed in previous work: the reconstruction is much
  faster when there is more energy concentrated in the known part
  \cite{fiddy83enforcing,fienup86phase}. Quite surprisingly, this can
  be explained using our previous results on phase retrieval with
  approximately known Fourier phase.

\end{abstract}

\section{Introduction}
\label{sec:introduction}

The phase retrieval problem amounts to signal reconstruction from the
magnitude of its Fourier transform. The problems arises in several
fields of physics including crystallography, astronomy, and certain branches
of optics and microscopy. As with any inverse problem
where the sought signal must be reconstructed from incomplete data,
the main questions are: uniqueness of the reconstruction and design of
an efficient reconstruction method that would be insensitive to
possible errors in the measurements. It it trivial to show that,
without additional information about the signal, the Fourier magnitude
alone cannot provide sufficient information for successful
reconstruction. Inasmuch as each Fourier domain component can be
assigned an arbitrary phase to accompany the measured magnitude
resulting in a signal that fully corresponds to the measurement,
namely the Fourier magnitude. One of the most widely used assumptions
is that the sought signal $z(m,n)$ has limited spacial extent, i.e.,
one can assume that outside the rectangle $[0,M-1]\times[0,N-1]$ the
signal values are zeros. In this case, it has been shown that in most
cases the phase retrieval problem has a unique solution up to a
trivial transformation. Possible
variations being: constant phase factor, shifts, and axis
reversal\cite{hayes82reconstruction,bruck79ambiguity}. Hence, in what
follows we shall not treat in depth the problem of uniqueness and will
concentrate on the efficient reconstruction method that provides good
reconstruction results even in the case of nosy measurements.

In this paper we develop an algorithm for signal reconstruction from
the magnitude of its Fourier transform in a situation where some
(non-zero) parts of the sought signal are known. Although our method
does not assume that the known part comprises the boundary of the
sought signal, this is often the case in microscopy: a specimen is
placed inside a known mask, which can be thought of as a known light
source that surrounds the unknown signal. Therefore, in the past,
several algorithms were suggested that solve the phase retrieval
problem assuming known boundary
values~\cite{hayes82importance,hayes83recursive,fienup83reconstruction,fienup86phase}. 
Unlike our method, these methods do rely on the fact that the known
part in on the boundary.

Besides the reconstruction method we give an explanation of the
phenomena observed in previous work: the reconstruction is much faster
when there is more energy concentrated in the known
part~\cite{fiddy83enforcing,fienup86phase}. Quite surprisingly, this
can be 
explained using our previous results on phase retrieval with
approximately known Fourier phase~\cite{osherovich11approximate}.

\section{Review of existing methods}
\label{sec:revi-exist-meth}
Before we proceed further it is important to note that the standard
Hybrid Input-Output method~\cite{fienup82phase} can be applied because
the Fourier magnitude of the sought signal is known. However, this
approach is not optimal as it does not use the additional information
available in this case. Furthermore, HIO fails in case of
complex-valued objects without tight support information.

Let us start with the notation used throughout the paper. As we
already mentioned, the sought signal is denoted by $z(m,n)$.  The signal is assumed to have
a finite support, specifically, it vanishes outside the box
$[0,M-1]\times[0,N-1]$, and our goal is, as before, to
reconstruct it from the (oversampled) magnitude of its Fourier
transform $|\hat{z}(p,q)|$, where
\begin{equation}
  \label{eq:boundary-1}
  \hat{z}(p,q) =
  |\hat{z}(p,q)|\exp(i\phi(p,q))\equiv\mathcal{F}[z(m,n)]\ .
\end{equation}
In our implementation, $F$ the \textit{unitary} Fourier transform. This
regularization constant is chosen with only one purpose: to make the
distance (norm) in the Fourier domain equal that in the object
domain. Thus, by the discrepancy (error) in the Fourier domain one can
easily estimate the error in the object domain. Hence,
\begin{equation}
  \label{eq:boundary-2}
  \mathcal{F}[z(m,n)]=(PQ)^{-1/2}\sum_{m=0}^{P-1}
  \sum_{n=0}^{Q-1}z(m,n)\exp[-i2\pi (mp/P + nq/Q)]\ ,
\end{equation}
where $m,p=0,1,\ldots,P-1$ and $n,q=0,1,\ldots,Q-1$. In addition, we
require the ``two-fold oversampling'' in the Fourier domain, that is
$P=2M-2$, $Q=2N-2$. The purpose of this oversampling is to capture the
information that $z(m,n)$ has only limited support. With these
definitions, there exists one well-known relation between the squared
magnitude of the Fourier transform and the linear (as opposed to
cyclic) auto-correlation (denoted by $\star$) function of $z(m,n)$:
\begin{equation}
  \label{eq:boundary-3}
  \begin{split}
    r(i,j)
    & \equiv z(m,n) \star z(m,n) \\
    & \equiv \sum_{m=0}^{M-1}\sum_{n=0}^{N-1}z(m,n)z^{*}(m-i,n-j)\\
    & = \mathcal{F}^{-1}(|\hat{z}(p,q)|^{2})\ ,
  \end{split}
\end{equation}
where $r(i,j)$ is of size $(2M-1)\times(2N-1)$ pixels defined over the
region $1-M\leq i \leq M-1$, $1-N\leq j \leq N-1$. This relation
between the magnitude of the Fourier transform and the
auto-correlation function of the sought signal was used by Hayes and
Quatieri to develop an elegant algorithm for finding
$x$ (note that $x$ is not necessarily equal to $z$ due to
  possible non-uniqueness) when its boundaries are
known~\cite{hayes83recursive,hayes82importance}. The algorithm assumes
that the boundaries, that is, the first row $z(0,:)$,\footnote{Here we
  use MATLAB notation, where a colon (:) denotes the entire vector of
  indexes of the corresponding dimension.} the last row $z(M-1,:)$,
as well as the first and the last columns: $z(:,1)$, and $z(:,N-1)$,
are known. The algorithm is iterative and at each iteration it
recovers two new unknown rows (or columns) of $x$. The authors
demonstrated that the $k$-th iteration of the algorithm reduces to a
simple matrix (pseudo) inverse to solve the following system of
equations
\begin{equation}
  \label{eq:boundary-4}
  \left[
    \begin{array}{c}
      F \ \vert\  L
    \end{array}
  \right] \,
  \left[
    \begin{array}{c}
      x(N-k, :)\\
      x(k, :)
    \end{array}
  \right] =
  \left[    
   \begin{array}{c}
      \tilde{r}(N-k, :)
    \end{array}
  \right] \,, 
\end{equation}
where the matrices $F$ and $L$ correspond to the cross-correlation
with the first and the last rows, respectively. The right-hand side
$\tilde{r}(N-k)$ is obtained from the $(N-k)$-th row of the
auto-correlation function $r(i,j)$ from which the contribution of
already recovered rows $1,2,\ldots,k-1$ and $M-2,M-3,\ldots, M-k+1$
has been subtracted. For a more detailed description we address the
reader to the original articles.  The most appealing property of this
algorithm is that it requires only a small, known in advance, number
of iterations until the whole signal $x(m,n)$ is recovered. The authors
also provided conditions that they claimed were sufficient to
guarantee uniqueness of the reconstruction. The latter, however, were
proven to be incorrect (see \cite{fienup86phase}). Nonetheless, the biggest
problem with this algorithm is not the non-uniqueness of
reconstruction, because, as we already mentioned, the two-dimensional phase
retrieval solution is usually unique (up to trivial transformations)
to start with. The main difficulty  
that makes the algorithm impractical for all but tiny problems is its
numerical instability. It can easily be shown that the error grows
exponentially due to the recurrent nature of the algorithm. 
Even if we assume that the measurements are ideal, containing
absolutely no error, each iteration of the algorithm will introduce a
small error due to the finite computer precision. In the next
iteration, the norm of this error will be increased by a factor
proportional to the condition number of the matrix
$[F\,\vert\,L]$. The new error will be further amplified (by the same
factor) in the next iteration, and so on. This will result in
extremely fast (exponential) error growth. This is demonstrated in
Figure~\ref{fig:hayes-quatieri-reconstruction} where a small
($128\times128$ pixels) image was reconstructed by the HQ algorithm in
the horizontal direction, that is, reconstructing column after
column. This exponential growth is observed whenever the condition
number of the matrix $[F\,\vert\,L]$ is greater than one. It can
equal unity in some very special cases, for example, when the known
boundaries contain a single delta function. This situation was
considered in \cite{fienup83reconstruction,fienup86phase}, and in
\cite{fiddy83enforcing}, although in the latter it was not used
directly for the reconstruction---the authors added this condition to
guarantee the uniqueness of the reconstruction. Moreover, all of them
observed empirically that the reconstruction was faster when the known
part contained more energy. This observation is common, even though
the authors use different reconstruction methods. None of them,
however, provided an explanation for this phenomenon. In the next
section we present our reconstruction routine and explain why a
``strong'' known part leads to a fast and stable reconstruction.
Additionally, we will consider the influence of noise in the
measurement---another issue that has been largely overlooked in 
previous works despite its enormous importance.

\section{Our reconstruction method}
\label{sec:our-reconstr-meth}
\begin{figure}[H]
  \centering
  \includegraphics[width=8cm]{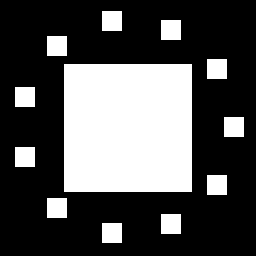}
  \caption{Artificial mask design.}
  \label{fig:mask}
\end{figure}

First, we must consider the source of the known boundary. In
microscopy, it is often natural to create a mask (for transparencies)
or a bed (for light reflecting objects) whose boundaries are known and
designed in a way that leads to easy image reconstruction. An example of
such a mask is shown in Figure~\ref{fig:mask}.

The object here is
assumed to be transparent, so the white areas correspond to simple
windows in an opaque material (shown in black). The big window in the
center is where the object is placed. The whole
construction is then illuminated by a coherent plane wave such that
the small windows in the mask can be assumed to be of known intensity.
There is nothing special about the plane wave here, the
most important point is that part of the  image is
known. This setup allows us to formulate the following minimization
problem to find the unknown signal $x(m,n)$:
\begin{equation}
  \label{eq:boundary-5}
  \begin{split}
    \min_{x} &\quad \||\mathcal{F}[x + b]| - r\|^{2}\\
    \mathrm{subject\ to} &\quad x(m\in \mathcal{O}_{m}, n\in
    \mathcal{O}_{n})=0 \,, 
  \end{split}
\end{equation}
where $r$ denotes the measured Fourier magnitude of the entire signal,
$b$ represents the known part (boundary) and $(\mathcal{O}_{m},
\mathcal{O}_{n})$ designate the 
location of the off-support parts of $x$ (basically, these are the
locations occupied by the mask except the central window where the
object is located). Note that we, again, use $x$ to denote the
reconstruction result because, in general, it may not be equal the
sought signal $z$.

Note tat the mask can, in principle, be constructed in a way that
makes the reconstruction trivial: it is sufficient to place only one
infinitesimally small window (a delta function) at a sufficient
distance from the object. In this case the mask would satisfy the
holography conditions and the reconstruction could be as easy as
applying a single Fourier transform.  However,
generating a delta function is not possible in
practice. Practical mask design must balance between the production
costs/difficulties and how helpful it is for the reconstruction
process. Hence, here we use simple square windows of relatively large
size, located close to the object. Hence, a non-iterative
reconstruction is not possible. However, using a quasi-Newton
optimization method to solve (\ref{eq:boundary-5}) we were able to get
good results, as demonstrated in Section~\ref{sec:numerical-results}.

\begin{figure}[H]
  \centering
  \includegraphics{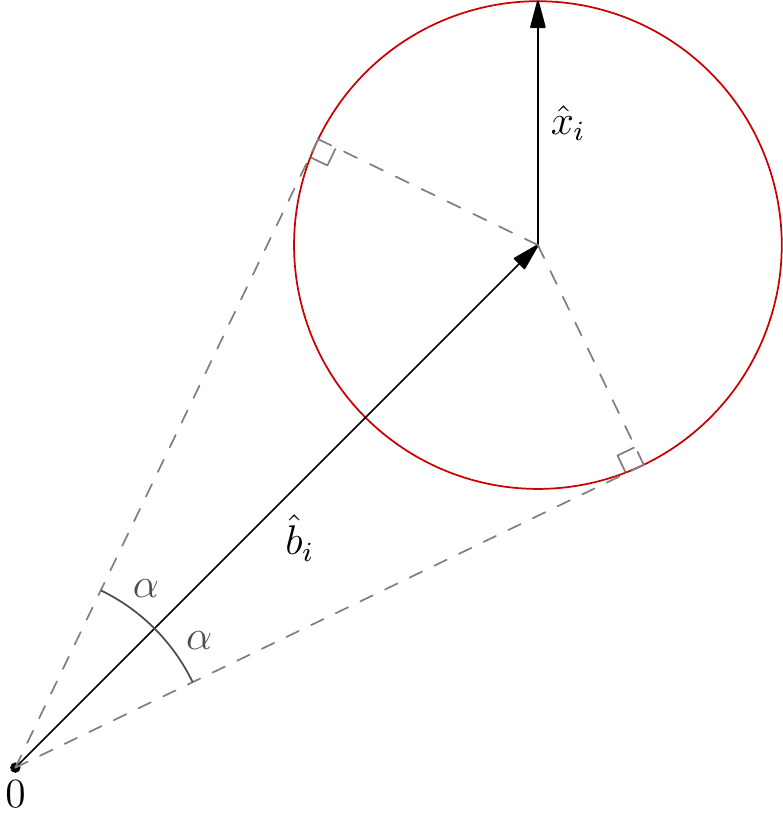}
  \caption{``Strong'' known part leads to the phase retrieval problem
    with approximately known Fourier phase.}
  \label{fig:strong-boundary}
\end{figure}

Before we proceed to the
numerical results, it is pertinent to discuss briefly the design of
the mask. It is not known at the moment what is the best way to design
the mask so that the reconstruction would be fast and robust. Our
experience shows that the mask should have a strong ``presence'' in
all frequencies. More precisely, we can explain why this situation
leads to a fast reconstruction. Consider a single element in the
Fourier space. The contribution of the known part (denoted by
$\hat{b}_{i}$) is fully determined, that is, its magnitude and phase
are known. Hence, the full Fourier domain data at this location is the
sum $\hat{x}_{i}+\hat{b}_{i}$ that is located somewhere on the red circle
shown in Figure~\ref{fig:strong-boundary}. Note that the phase
$\angle(\hat{b}_{i} + \hat{x}_{i})$ is known up to $\pi/2$ radians
when $\alpha \leq \pi/4$ (see Figure~\ref{fig:strong-boundary}). That
is, when
\begin{equation}
  \label{eq:boundary-6}
  \arcsin
  \left (
    \frac{|\hat{x}_{i}|}{|\hat{b}_{i}|}
  \right ) \leq \frac{\pi}{4}
  \Longleftrightarrow |\hat{b}_{i}| \geq \sqrt{2} |\hat{x}_{i}| \,. 
\end{equation}

The relation in Equation~\eqref{eq:boundary-6} is quite interesting:
if the energy concentrated in the known part is at least twice as
large as the energy in the unknown part\footnote{The energy must also
  be distributed ``well'' in the frequency domain.}, the problem
reduces to phase retrieval with Fourier phase known within the limit
of $\pi/2$ radians. Hence, according to our results
in~\cite{osherovich11approximate}, \emph{any} reasonable algorithm will be 
able to reconstruct the unknown part. Moreover, the larger the ratio
$|\hat{b}_{i}|/|\hat{x}_{i}|$, the smaller the Fourier phase
uncertainty is.  However, there is another aspect that is important in
practice---in all physical experiments, the measurements
inevitably contain some noise. In our case, where the measurements are
light intensity, the noise is well approximated by the Poisson
distribution. This means that stronger intensity will result in more
noise (though, the signal to noise ratio (SNR) usually increases with
the intensity growth). Hence, making the known part too strong
compared with the sought signal will make the measurement noise more
dominant with the respect to the unknown part and will eventually
arrive at the level where the reconstruction is not possible. Hence,
one should not just increase the intensity of the known part, as this leads
to poor reconstruction quality in case of noisy
measurements\footnote{Here we assume that the noise grows with the
  signal, as indeed happens with Poissonian noise.}. The most
appealing approach would be to design a mask that would use the
limited power in an efficient way. Without any a priori knowledge
about the sought signal, it seems that the optimal way would be to
create a mask whose Fourier domain power is spread evenly over all
frequencies. Note the relation to the holographic reconstruction where
a single small window (delta function) is used, because the Fourier domain
power of a delta function is the same over all frequencies. Instead of
using a delta function we can obtain a very good approximation to the
uniform power spectrum if some randomness is added to the mask
windows. It can be random shape of the windows or random
values/phases across them. Making random shapes may be more difficult
than adding a diffuser into a square window. Hence, in
Section~\ref{sec:numerical-results} we demonstrate the reconstruction
with mask windows of constant intensity and random intensity. As
expected, adding randomness improves the reconstruction speed and
noise-immunity.

\section{Numerical results}
\label{sec:numerical-results}
\begin{figure}[H]
  \centering
  \subfloat[]{\includegraphics{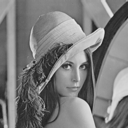}}
  \qquad{}
  \subfloat[]{\includegraphics{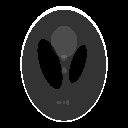}}
  \caption[Test images]{Test images (intensity).}
  \label{fig:boundary-test-images}
\end{figure}

\begin{figure}[H]
  \centering
   \subfloat[]{
    \includegraphics[width=0.45\textwidth{}]{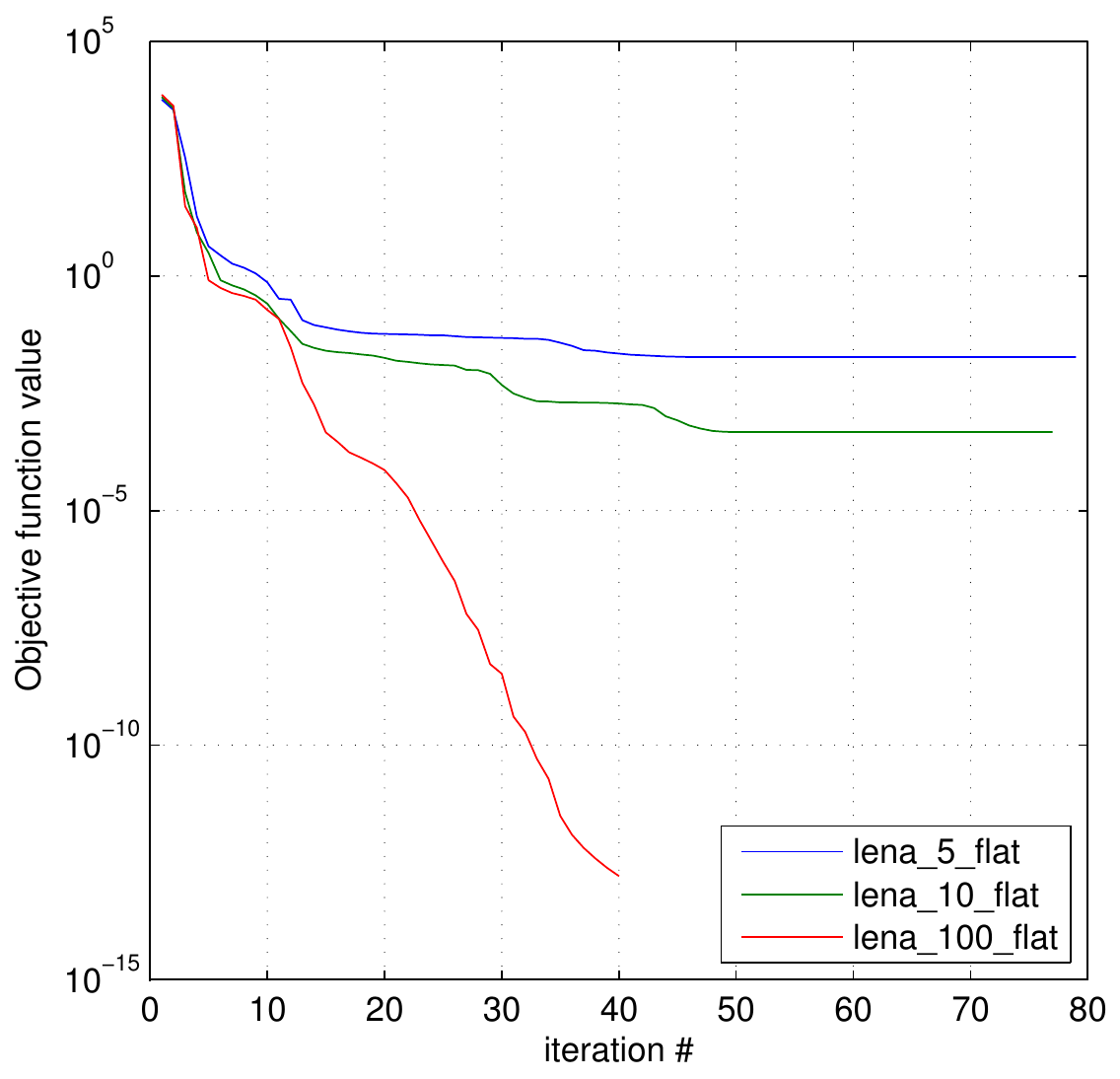}
  }
  \quad{}
  \subfloat[]{
    \includegraphics[width=0.45\textwidth{}]{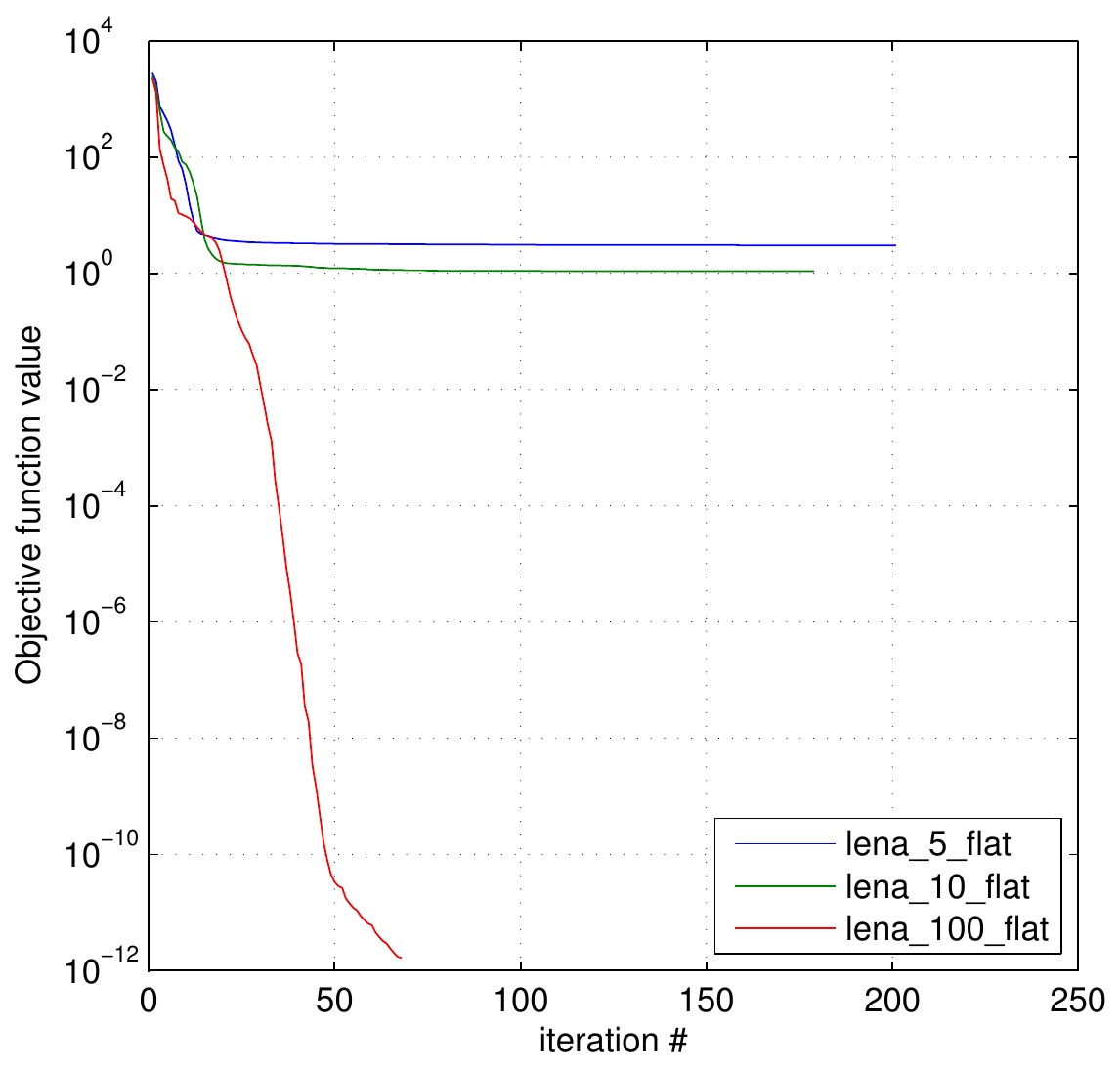}
  }
  \caption[Lena's reconstruction speed with flat mask
  values.]{Lena's reconstruction speed with flat mask values: (a)
    real-valued, (b) complex valued.}
  \label{fig:lena-reconstruction-speed-flat-mask}
\end{figure}

\begin{figure}[H]
  \centering
   \subfloat[]{
    \includegraphics[width=0.45\textwidth{}]{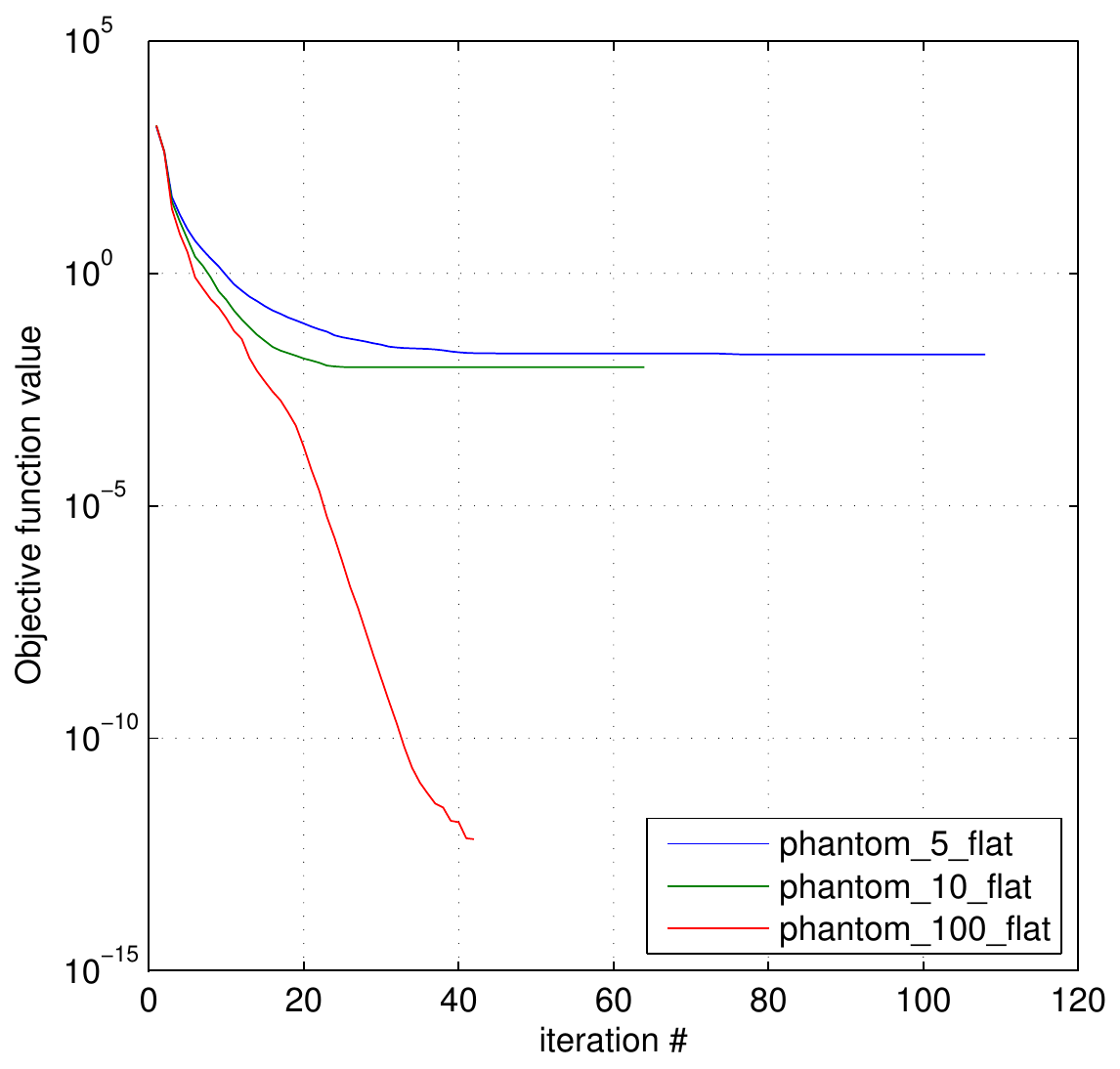}
  }
  \quad{}
  \subfloat[]{
    \includegraphics[width=0.45\textwidth{}]{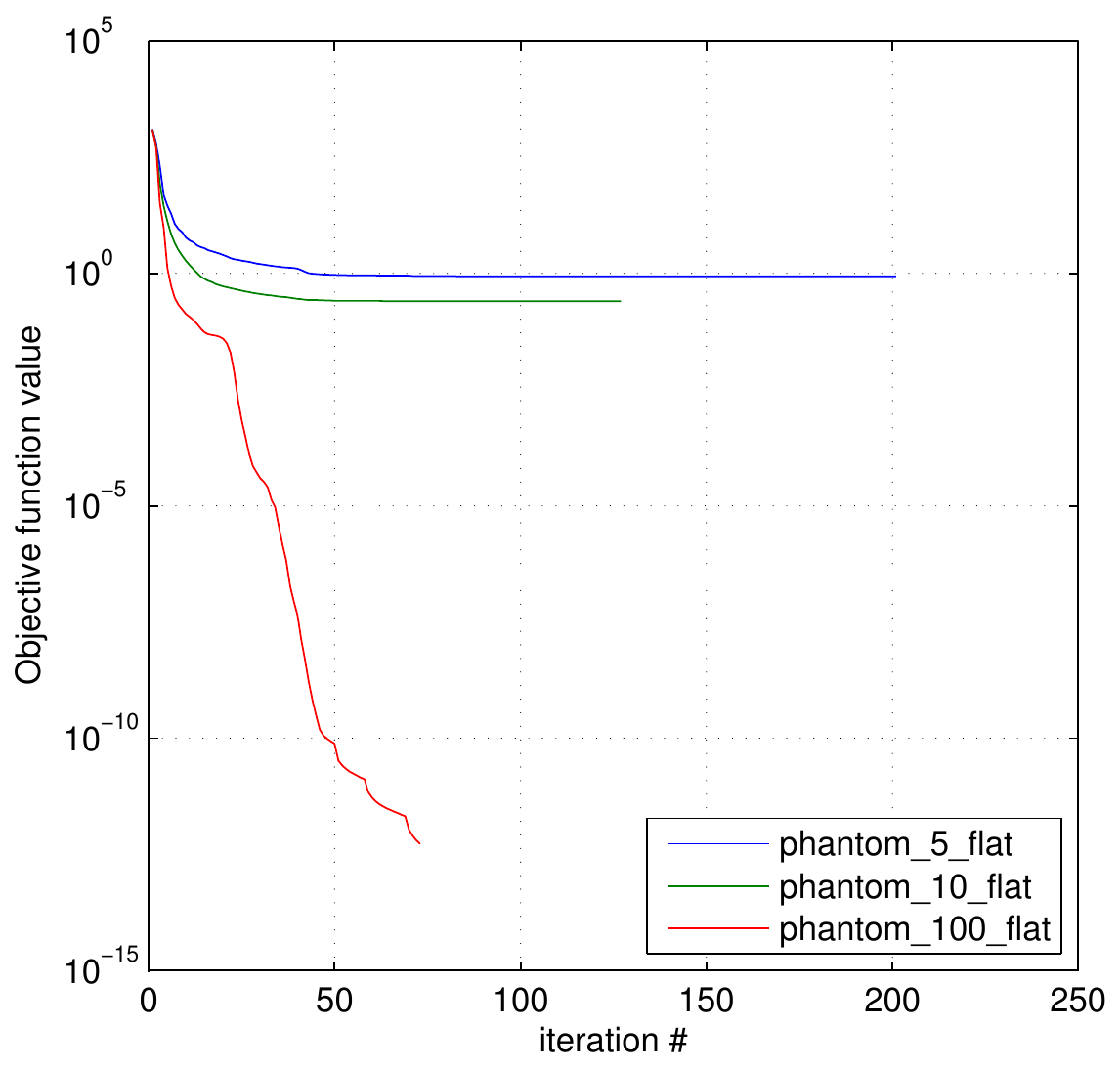}
  }
  \caption[Phantom's reconstruction speed with flat mask
  values.]{Phantom's reconstruction speed with flat mask values: (a)
    real-valued, (b) complex valued.}
  \label{fig:phantom-reconstruction-speed-flat-mask}
\end{figure}

We experimented with several images, however, the results provided
below are limited to two images of size $128\times 128$ pixels. These
were chosen to represent two different classes of objects. One is a
natural image ``Lena''  already used in our previous
experiments. Another is the Shepp-Logan phantom.  These images are
very popular in other fields. ``Lena'' is a classical benchmark in the
image processing community, because it has a lot of features and
delicate details. The second image, ``phantom'', is often used as a
benchmark in MRI related algorithms. However, its piece-wise constant
nature can approximate well objects that are often investigated in
microscopy, for example, cells. Besides the above differences these two
images differ by their support. Lena's support is tight, that is, it
occupies all the space and no shifts are possible. The phantom, on the
other hand, has non-tight support. As we saw earlier, this is an
important property for phase retrieval algorithms---complex valued
images with non-tight support are much more difficult for the current
reconstruction methods like HIO. The image intensities (squared
magnitude) is shown in Figure~\ref{fig:boundary-test-images}.
The images were tested in two different scenarios: first, they were
assumed to be real-valued and non-negative; second, they were made
complex-valued by adding a phase distribution. Here we present the
results for the case where the objects' phases were chosen to be
proportional to their intensity (scaled to the interval
$[0,2\pi]$). This choice corresponds to the case where the phase
changes in a relatively smooth manner.  However, all our experiments
indicate that the particular phase distribution has little effect on
the reconstruction, except the case where the object can be assumed
real non-negative. In this case the non-negativity prior can be used
to speed-up the reconstruction and improve its quality in the case of
noisy measurements, as is evident from
Figures~\ref{fig:lena-reconstruction-speed-flat-mask},
and~\ref{fig:phantom-reconstruction-speed-flat-mask}. 

\begin{figure}[H]
  \centering
  \subfloat[]{
    \includegraphics{lena_128}
  }\qquad{}
  \subfloat[]{
    \label{fig:hayes-quatieri-rec-intensity}
    \includegraphics{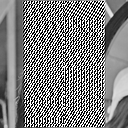}
  }\\
  \subfloat[]{
    \label{fig:hayes-quatieri-rec-error}
    \includegraphics[]{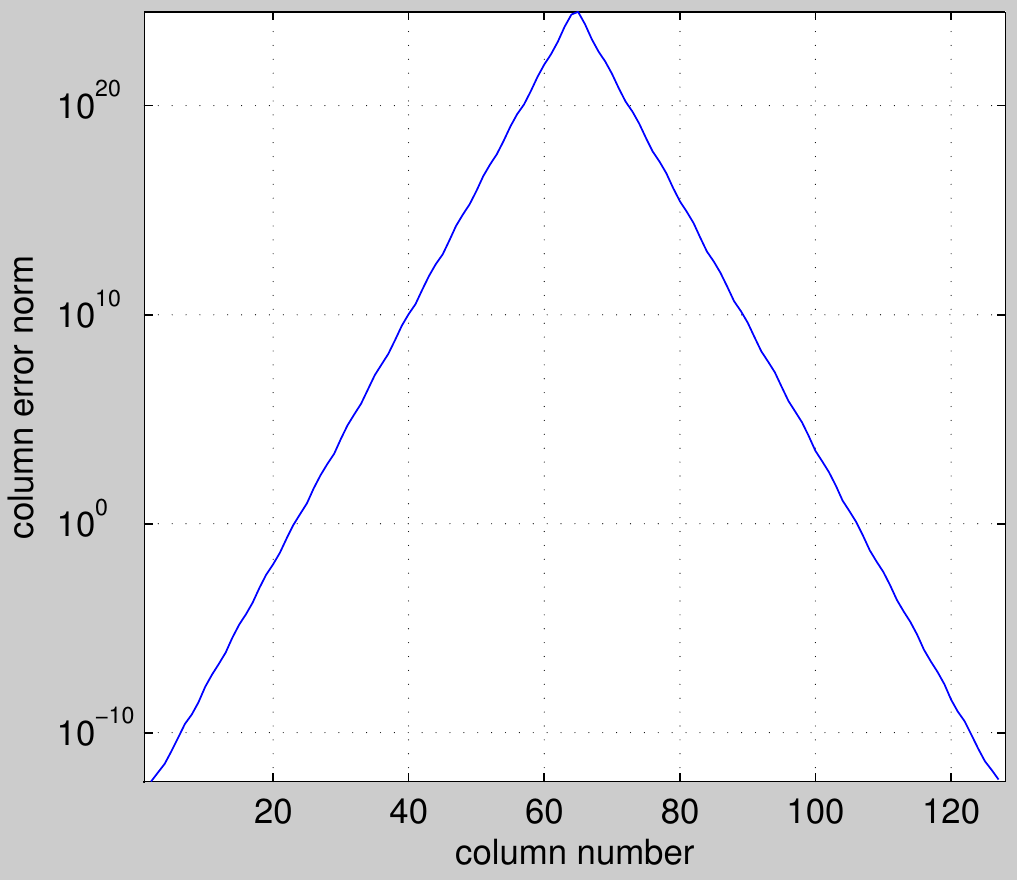}
  }
  \caption[Error growth in the Hayes-Quatieri recursive algorithm]{Error growth in the Hayes-Quatieri recursive algorithm:
  (a) original image, (b) reconstructed by the HQ algorithm, and (c)
  actual error growth in the HQ algorithm.}
  \label{fig:hayes-quatieri-reconstruction}
\end{figure}

Let us begin with a demonstration of the HQ algorithm
results. Following our discussion in Section~\ref{sec:revi-exist-meth}
we expect the error to grow exponentially as we progress from the
boundaries toward the image center. This expectation is confirmed by
the results shown in Figure~\ref{fig:hayes-quatieri-reconstruction}. Note
that the reconstructed intensity in
Figure~\ref{fig:hayes-quatieri-rec-intensity} cannot convey the true
error because the image storage format clips all values outside the interval
$[0,1]$. The true error is evident from
Figure~\ref{fig:hayes-quatieri-rec-error}.

\begin{figure}[H]
  \centering
  \subfloat[]{
    \includegraphics[width=8cm]{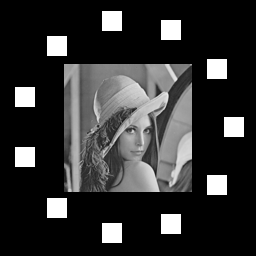}
  }
  \\
  \subfloat[]{
    \includegraphics[width=8cm]{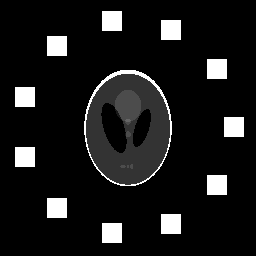}
  }
  \caption{Artificial mask with test images.}
  \label{fig:mask-withimages}
\end{figure}

In theory, we can use the same boundary conditions as the HQ
algorithm, however, our experiments indicate that the optimization
routine used in our method is prone to stagnation when the known
boundary (or image part in general) carries little energy. This is, of
course, in agreement with our discussion in the previous section: when
the boundary caries little energy it does not provide enough
information about the Fourier phase. This, in turn, causes line-search
optimization algorithms to stagnate
(see~\cite{osherovich11approximate}). The stagnation can be
alleviated by designing a mask that makes the reconstruction much
easier. In our approach the emphasis is also put on the simplicity of the
mask design. Hence, we used the simple mask shown in
Figure~\ref{fig:mask}.  The mask is of size $256\times 256$ pixels
with 11 square windows of size $20\times 20$ located approximately on
a circle of radius 100 pixels (see Figure~\ref{fig:mask}). The objects
are placed in the middle of the mask where a special window of size
$128\times128$ pixels is provided. The area outside this special
window comprises the known boundary. When placed in the mask, the test
images look as shown in Figure~\ref{fig:mask-withimages}.

Note also that the algorithm based on the minimization problem defined
in Equation~\eqref{eq:boundary-5} is very naive and does
not try to use the ``approximately known Fourier phase'' in the case
were the energy in the known part is sufficiently large. However, our
main goal is to show the influence of the mask design. This influence
is essentially independent of the algorithm used for reconstruction. 

\begin{figure}[H]
  \centering
   \subfloat[]{
     \includegraphics[width=0.45\textwidth{}]{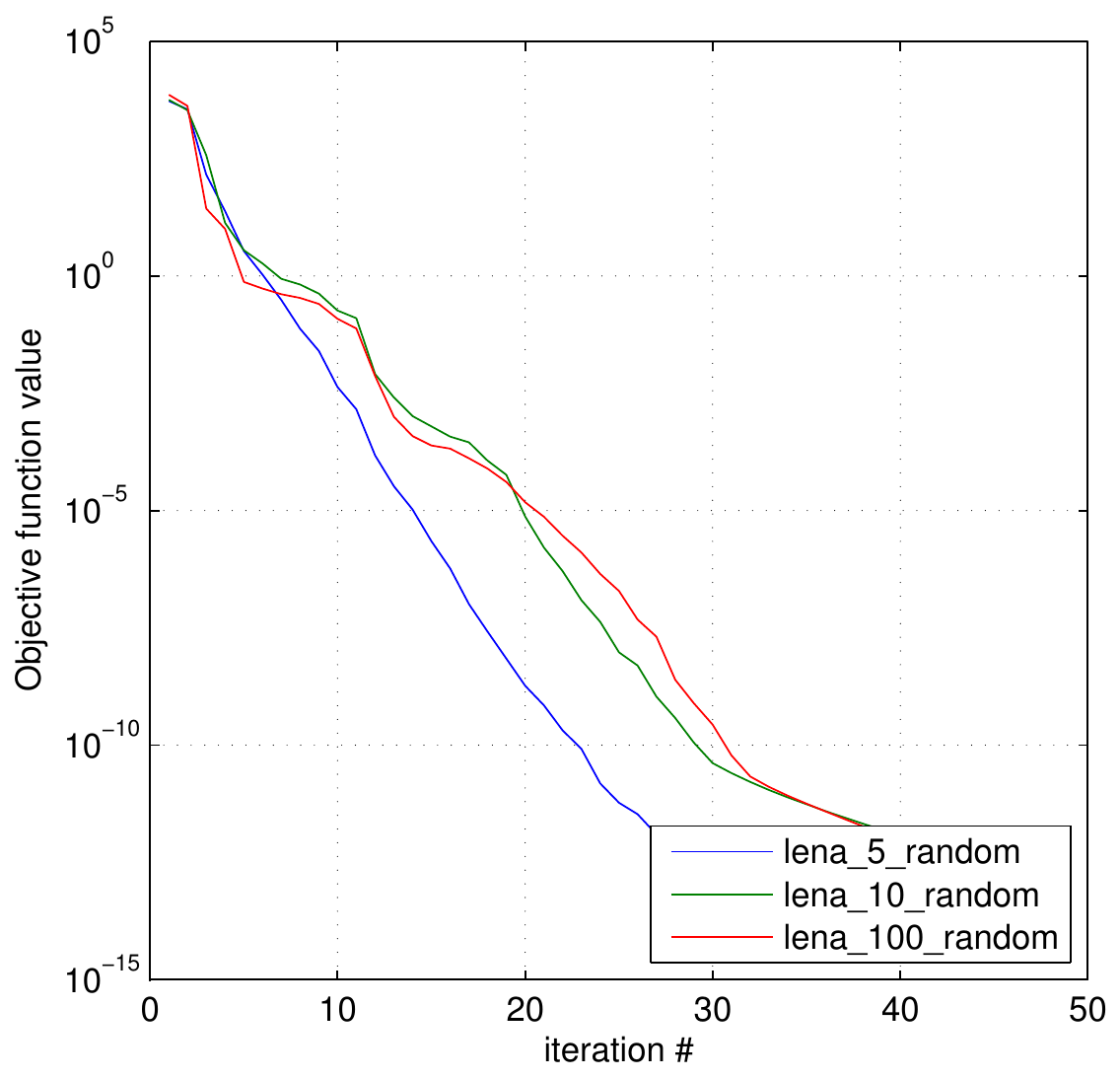}
  }
  \quad
  \subfloat[]{
    \includegraphics[width=0.45\textwidth{}]{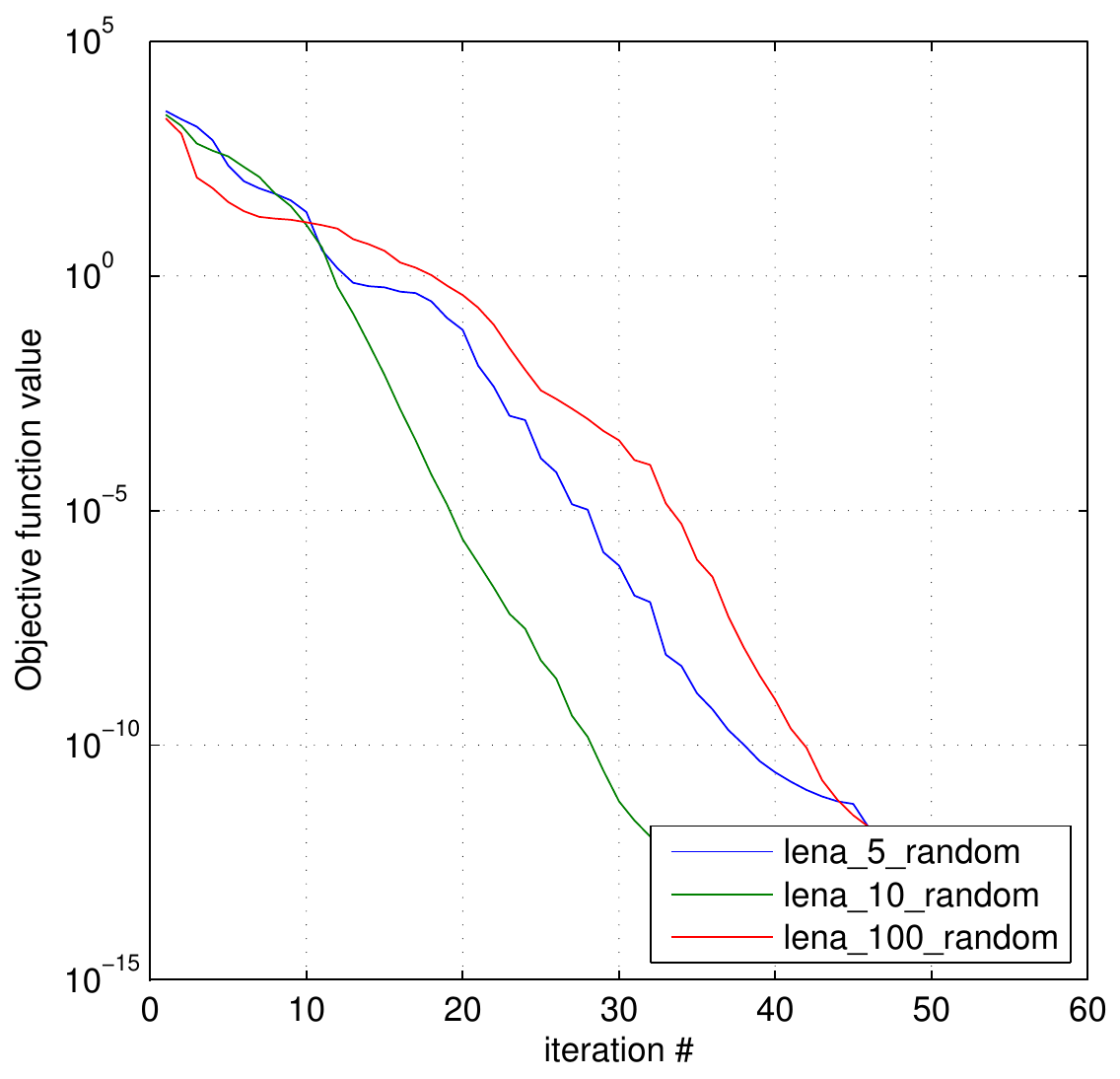}
  }
  \caption[Lena's reconstruction speed with random mask values]{Lena's
    reconstruction speed with random mask values: (a) real-valued, (b)
  complex-valued.}
  \label{fig:lena-reconstruction-speed-random-mask}
\end{figure}

\begin{figure}[H]
  \centering
   \subfloat[]{
     \includegraphics[width=0.45\textwidth{}]{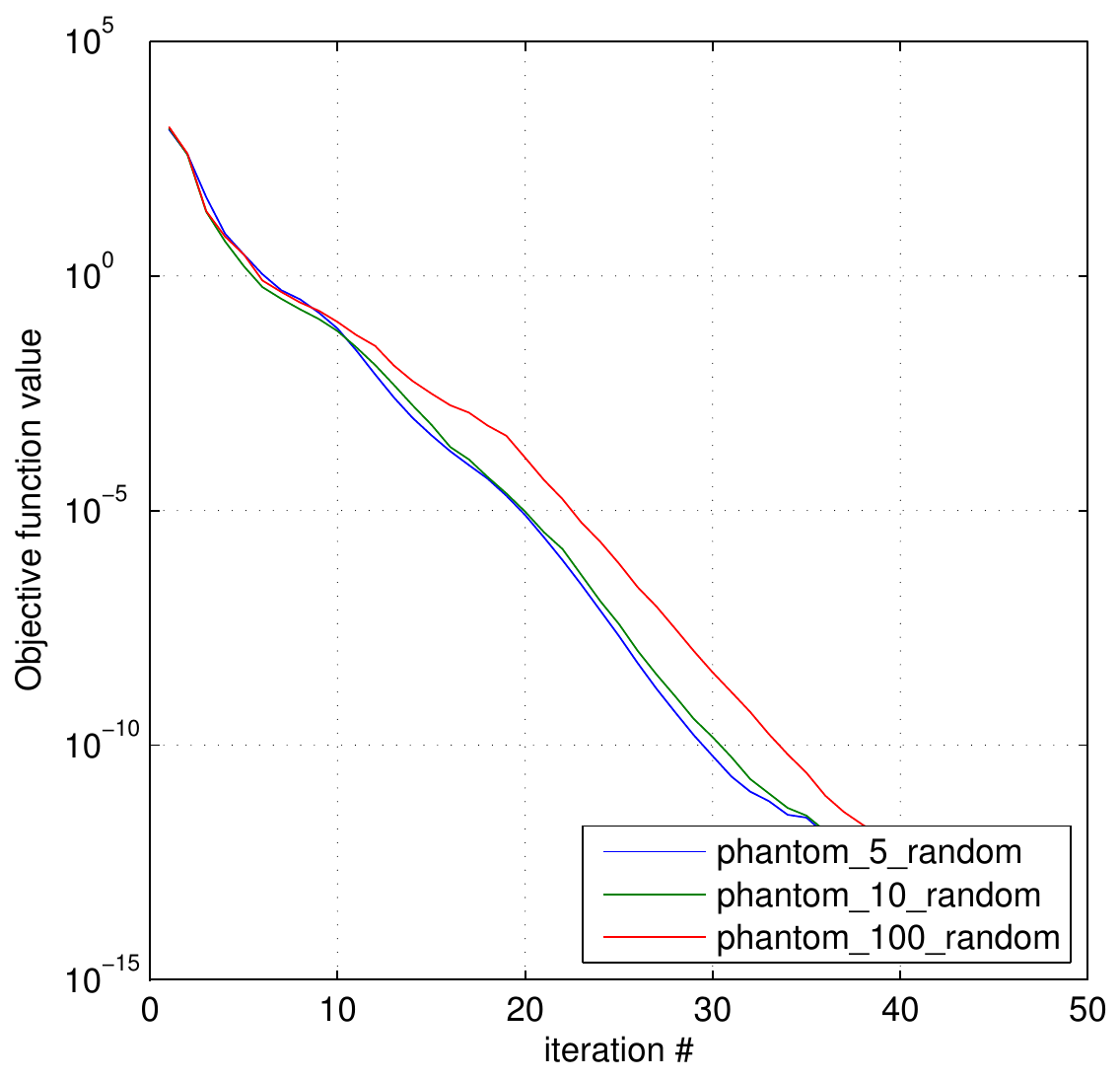}
  }
  \quad
  \subfloat[]{
    \includegraphics[width=0.45\textwidth{}]{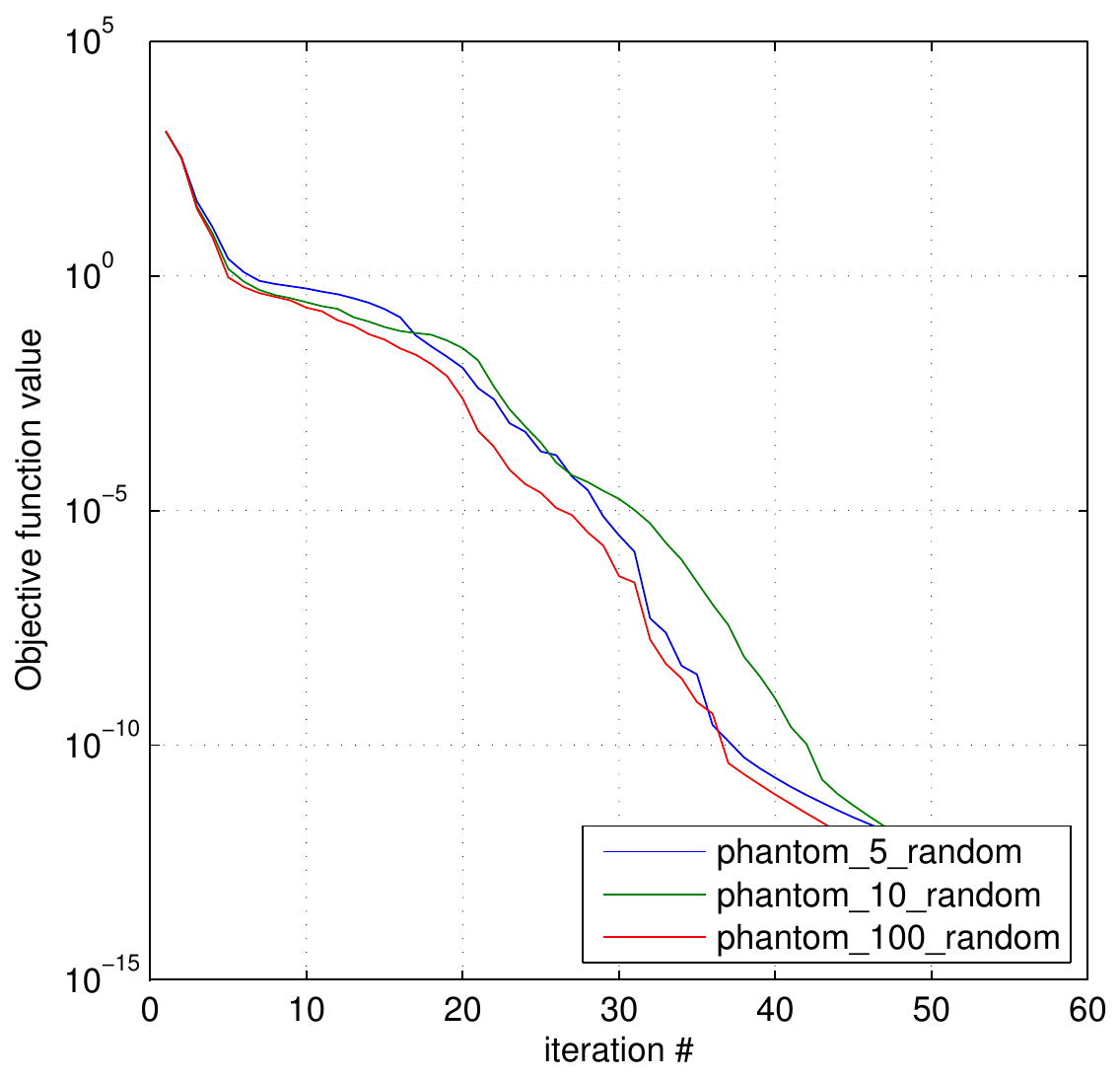}
  }
  \caption[Phantom's reconstruction speed with random mask
  values]{Phantom's reconstruction speed with random mask values: (a)
    real-valued, (b) complex-valued.}
  \label{fig:phantom-reconstruction-speed-random-mask}
\end{figure}

Let us demonstrate how the energy contained in the known part affects
the reconstruction. Recall that our expectation is that a mask that
has a strong ``presence'' in all Fourier frequencies will better suit
the reconstruction process in the noise-less case. The results shown
in Figures~\ref{fig:lena-reconstruction-speed-flat-mask},
and~\ref{fig:phantom-reconstruction-speed-flat-mask} fully support
this conjecture. These figures present the reconstruction speed (the
objective function minimization rate) for three different magnitudes
of the mask's windows values: 5, 10, and 100. The lowest value (5) was
chosen so as to be close to the theoretical ratio between the energy in
the known and the unknown parts that is required for guaranteed
reconstruction, as defined in Equation~\eqref{eq:boundary-6}. However, as
we can see the value of 5, and even the value of 10 did not result in
perfect reconstruction. This phenomenon has two reasons: first, 
doubling the amount of energy in the known part compared to the unknown part
is not sufficient---this energy must be distributed properly in the
Fourier domain; second, it may be attributed to the simplicity of the
chosen reconstruction algorithm. However, the second reason is less
likely in view of the following experiment. In an attempt to create a
``better'' distribution of the mask's energy in the Fourier domain we
added some ``randomness'' to the mask by modulating (multiplying) the
flat values across the mask's windows with random values in the
interval $[0,1]$. This step improved the speed of the reconstruction
and its robustness, as is evident from 
Figures~\ref{fig:lena-reconstruction-speed-random-mask},
and~\ref{fig:phantom-reconstruction-speed-random-mask}.
``Random'' mask are good for signals without any estimate of the
Fourier magnitude, on the other hand, when the Fourier magnitude of
the sought signal is known precisely, the ideal mask would have the
same magnitude multiplied by two. Hence, the ideal mask would be the
sought signal multiplied by two. This is, of course, is not a
practical situation. In practice, however, one may know approximately
the Fourier magnitude of the sought signals, in this case the best
mask can be obtained by another run of a phase retrieval method in
order to match the Fourier magnitude of the sought signal.

\begin{figure}[H]
  \centering{}
  \subfloat[]{
    \includegraphics[width=0.45\textwidth{}]{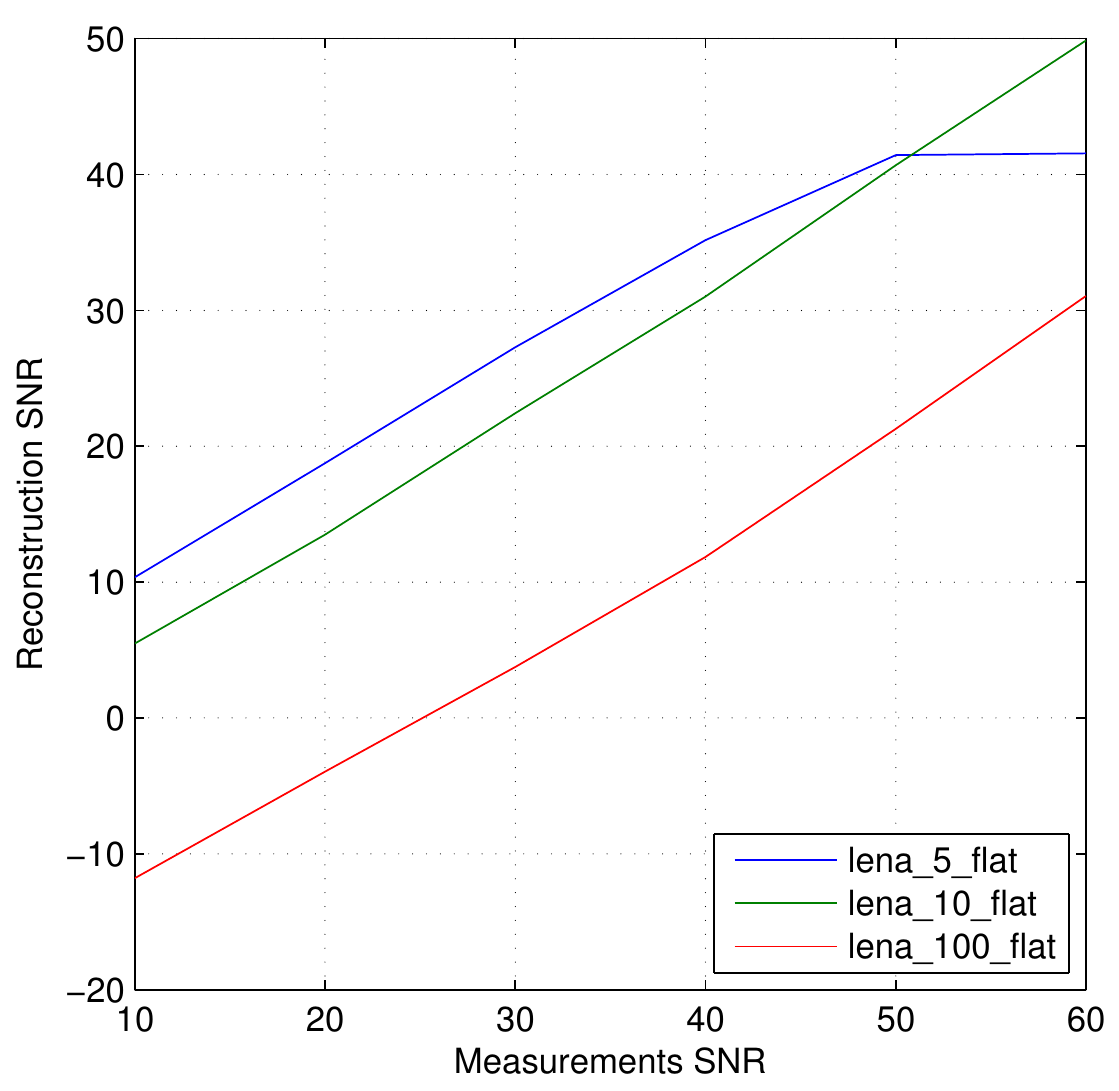}
  }
  \quad{}
  \subfloat[]{
    \includegraphics[width=0.45\textwidth{}]{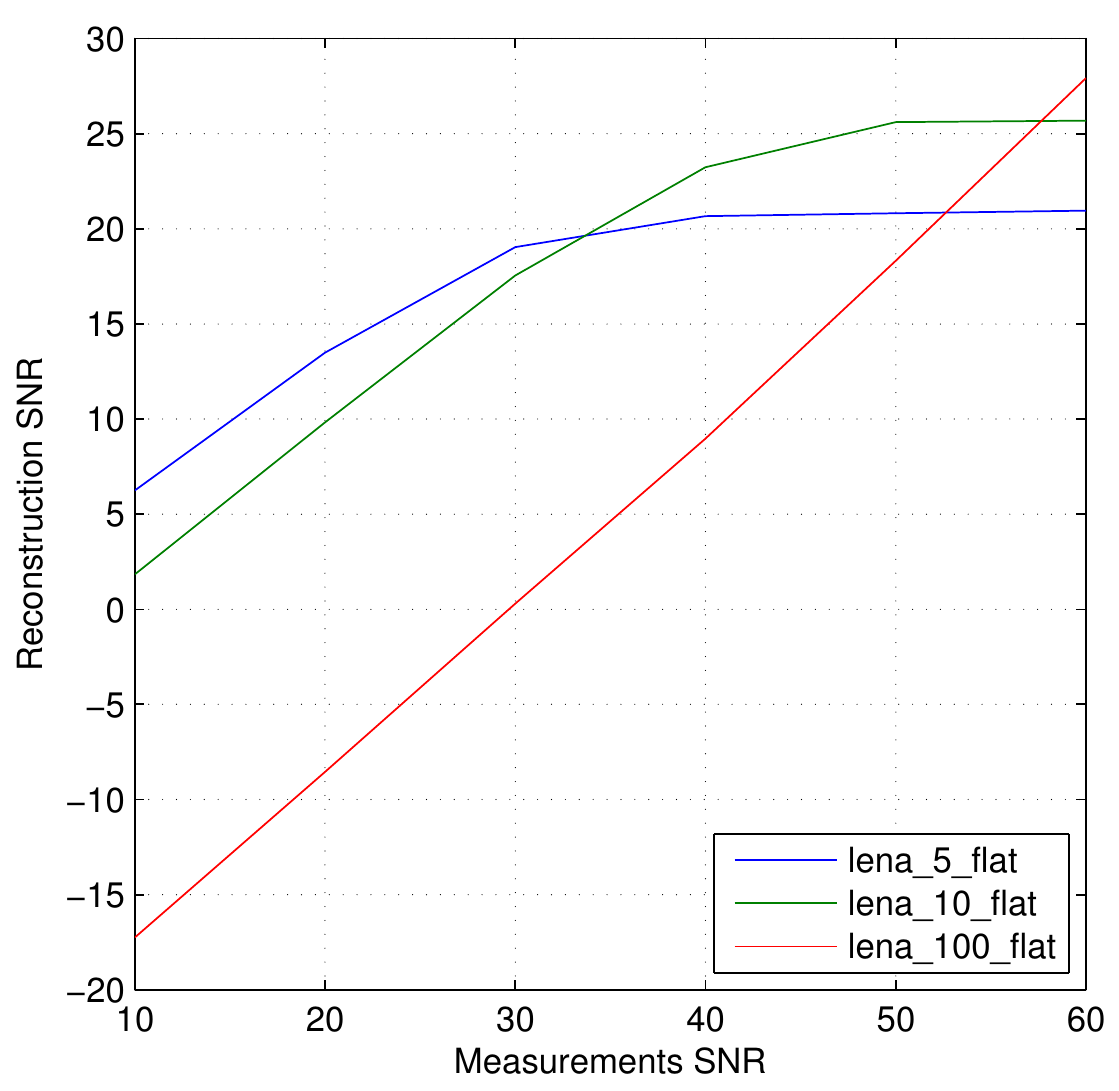}
  }
  \\
  \subfloat[]{
    \includegraphics[width=0.45\textwidth{}]{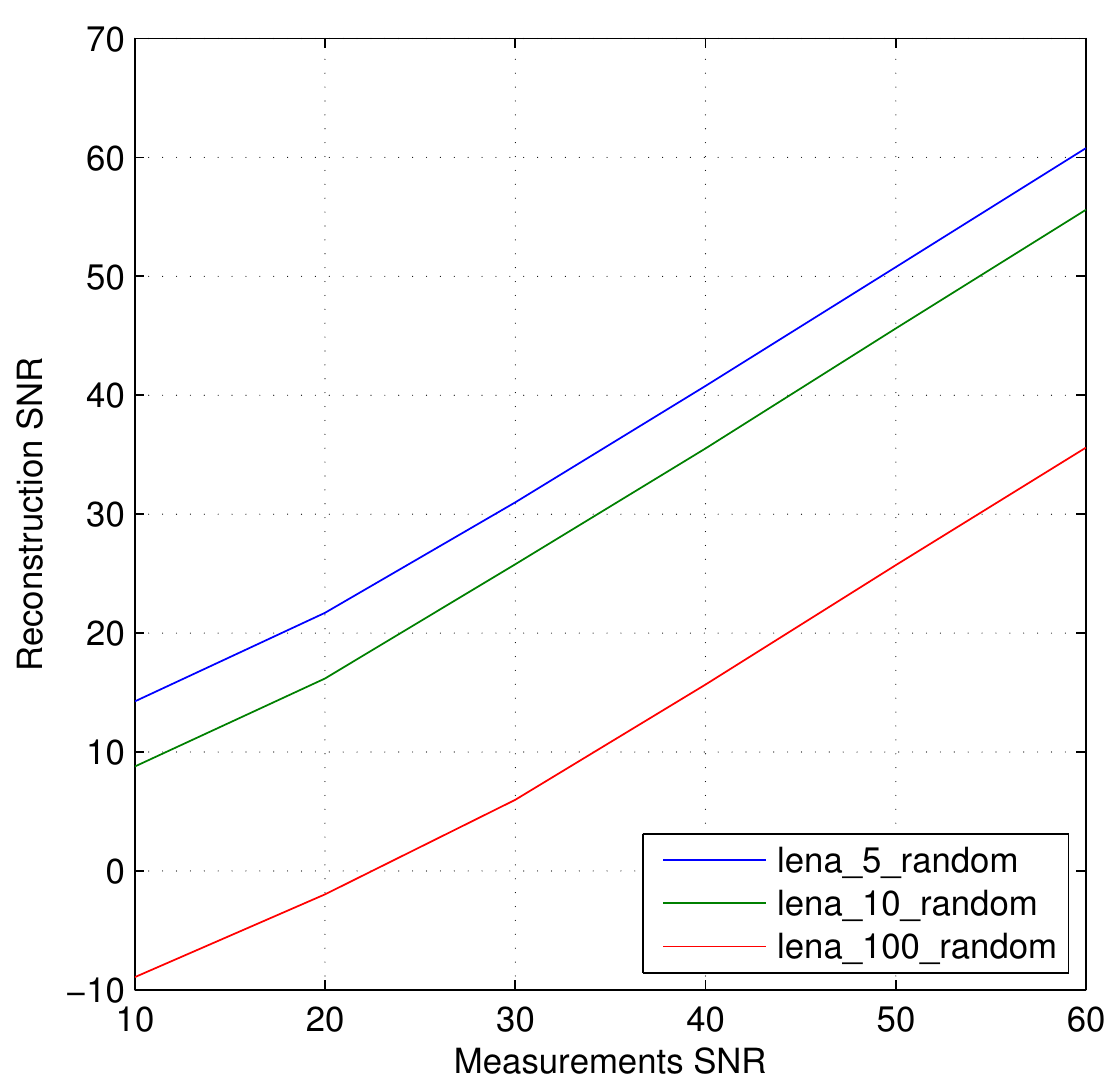}
  }
  \quad{}
  \subfloat[]{
    \includegraphics[width=0.45\textwidth{}]{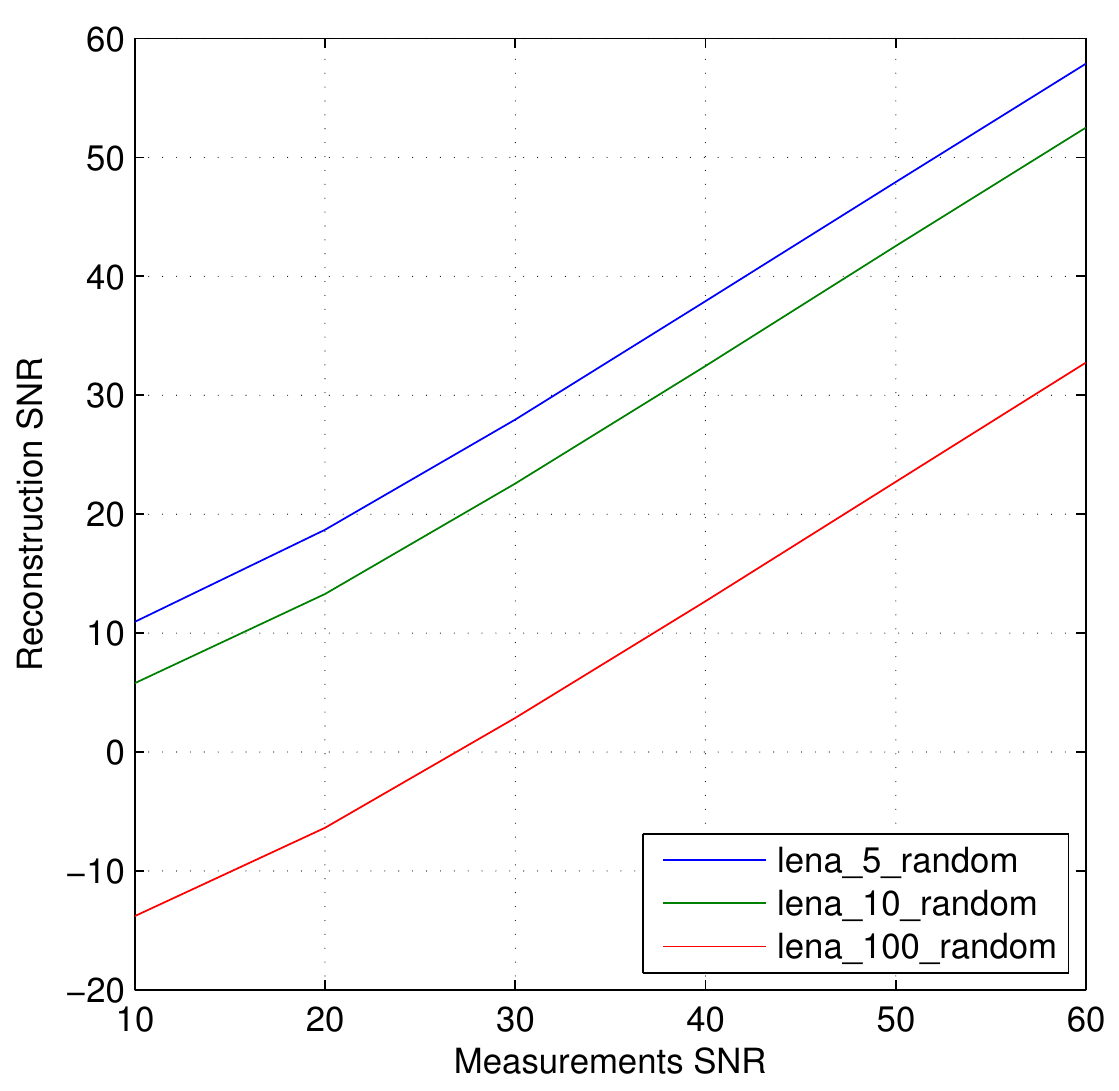}
  }
  \caption[Lena's reconstruction quality]{Lena's reconstruction
    quality: (a) real-valued with flat mask value, (b) complex-valued
    with flat mask value, (c) real-valued with random mask values, (d)
    complex-valued with random mask values.}
  \label{fig:lena-reconstruction-quality}
\end{figure}

\begin{figure}[H]
  \centering{}
  \subfloat[]{
    \includegraphics[width=0.45\textwidth{}]{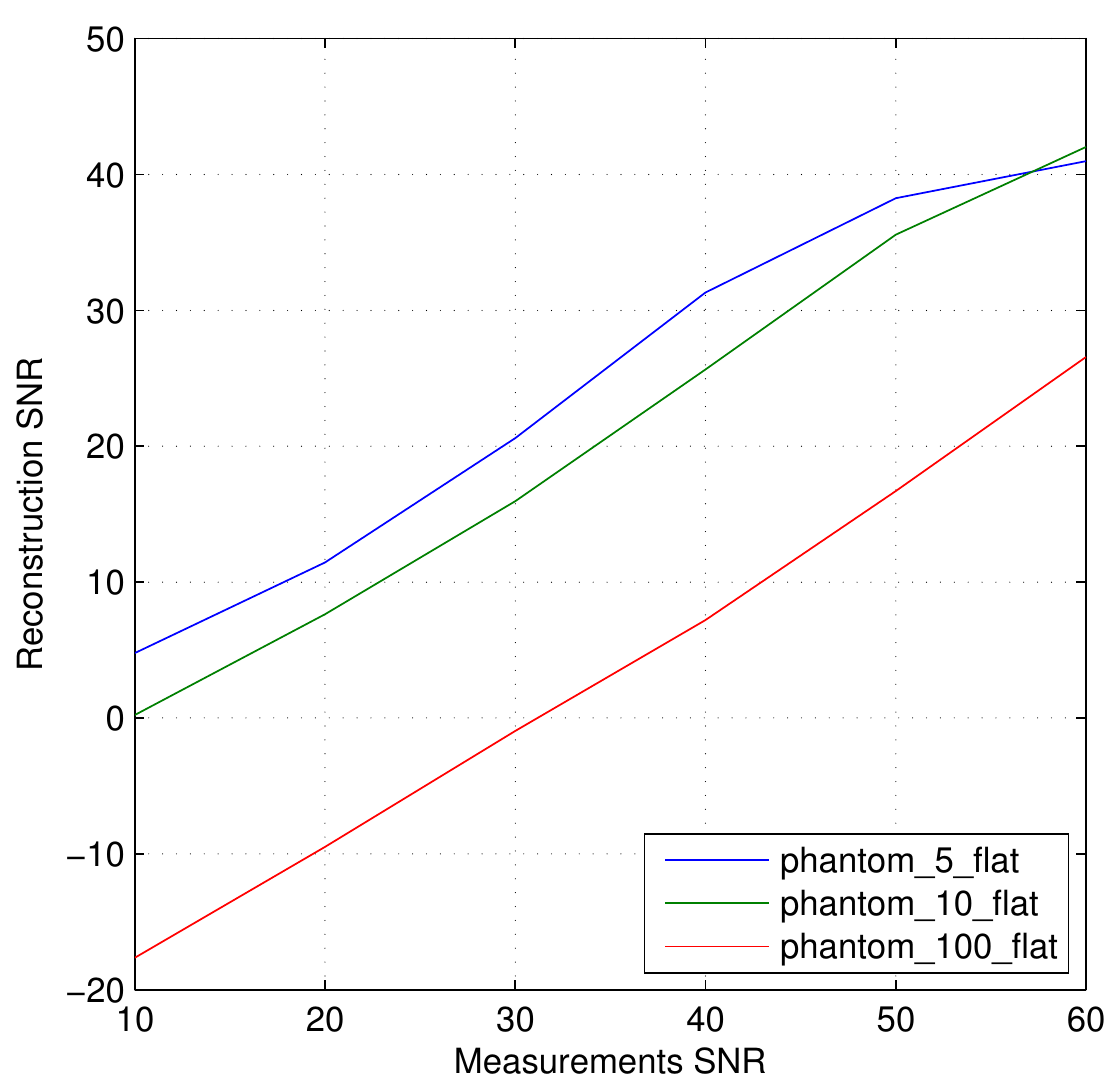}
  }
  \quad{}
  \subfloat[]{
    \includegraphics[width=0.45\textwidth{}]{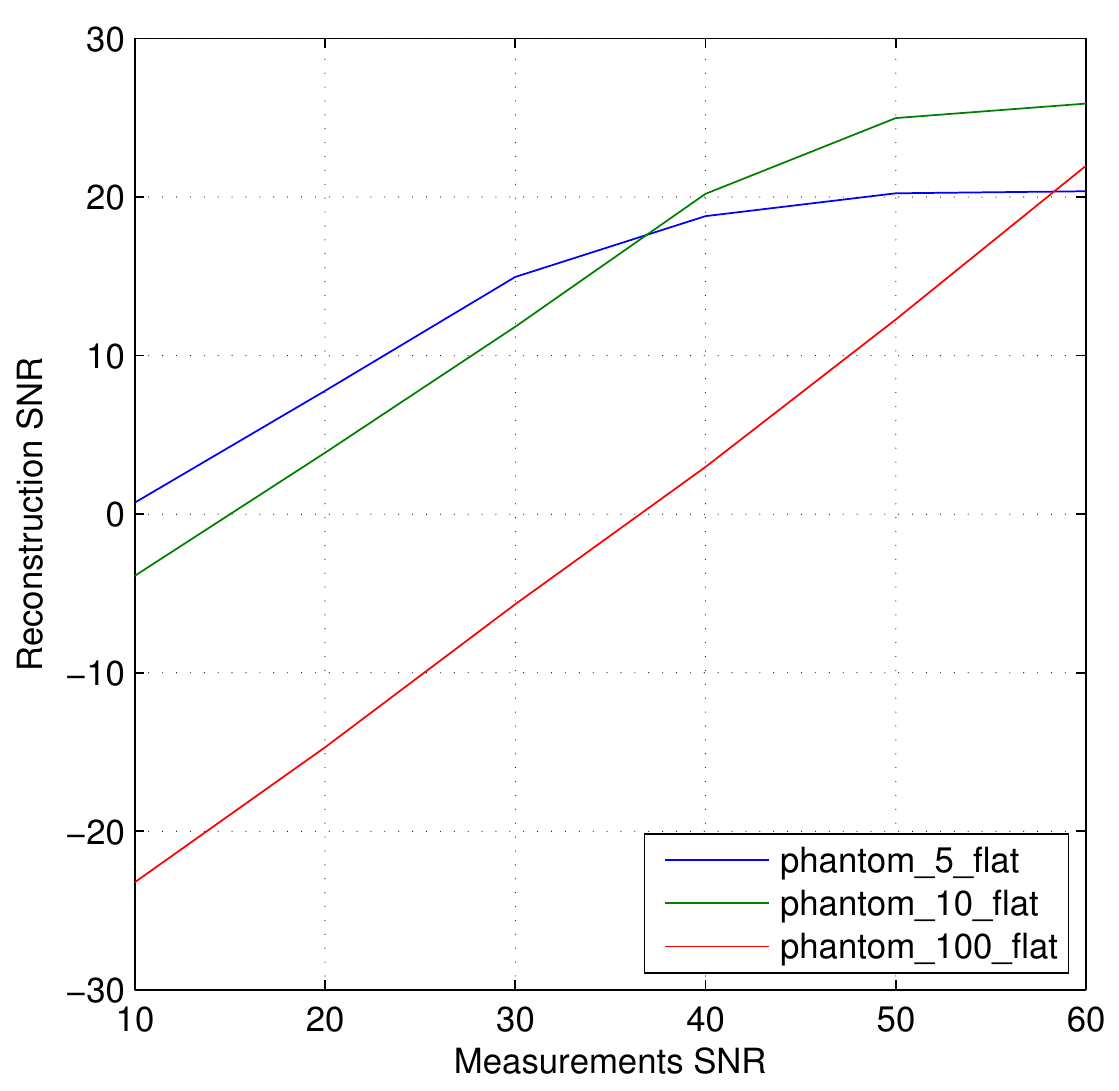}
  }
  \\
  \subfloat[]{
    \includegraphics[width=0.45\textwidth{}]{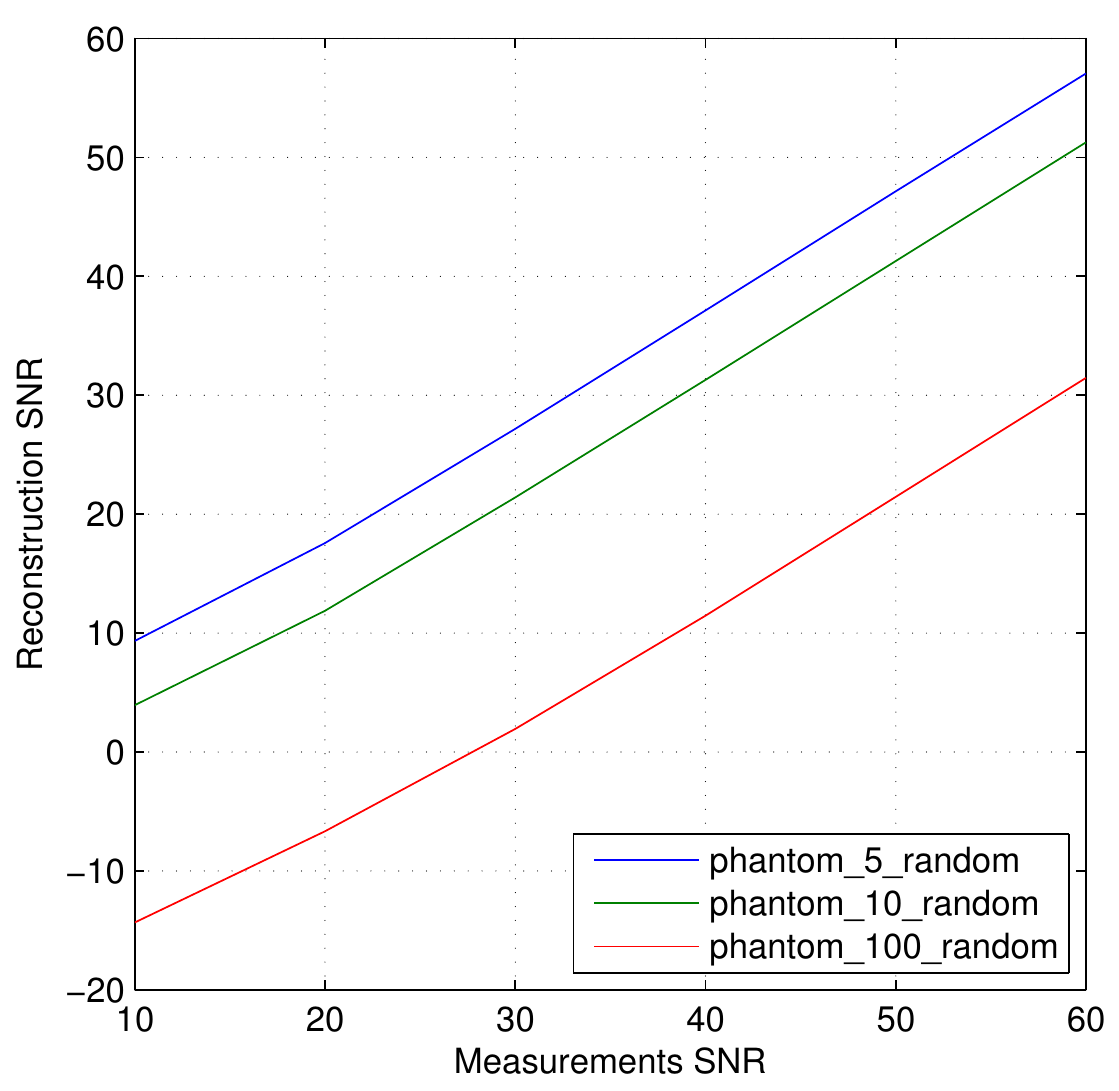}
  }
  \quad{}
  \subfloat[]{
    \includegraphics[width=0.45\textwidth{}]{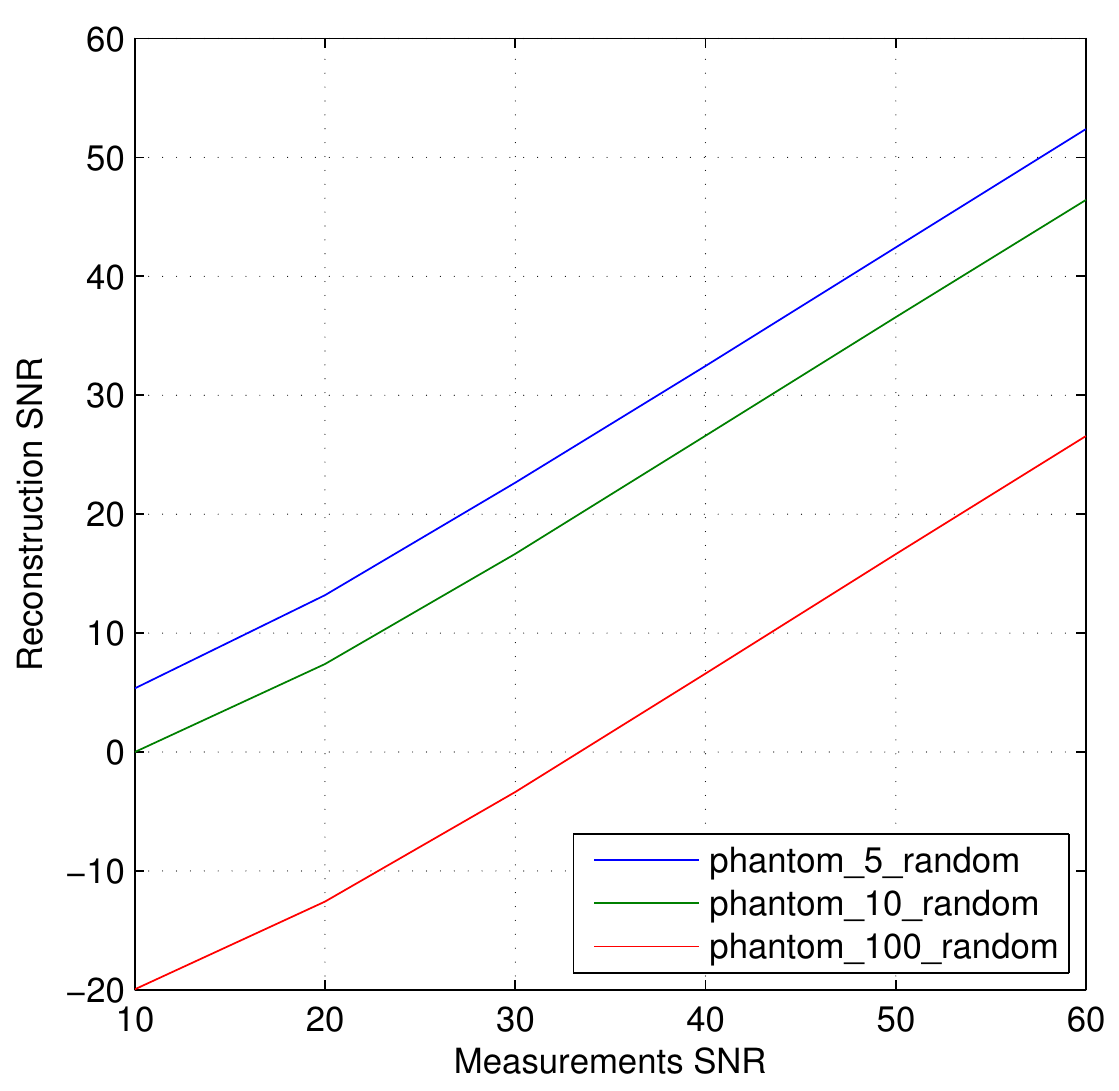}
  }
  \caption[Phantom's reconstruction quality]{Phantom's reconstruction
    quality: (a) real-valued with flat mask value, (b) complex-valued
    with flat mask value, (c) real-valued with random mask values, (d)
    complex-valued with random mask values.}
  \label{fig:phantom-reconstruction-quality}
\end{figure}

Note that now the reconstruction is successful for all mask values and
is very fast. However, despite this fast convergence, one must be
careful not to put too much energy into the known part. This approach
may harm the reconstruction quality of the unknown part when there is
noise in the measurements, as we demonstrate next.  In these
experiments the measurements (intensity values) were contaminated with
Poisson noise with different SNR ranging from 10 to 60 decibels. As is
evident from Figures~\ref{fig:lena-reconstruction-quality}
and~\ref{fig:phantom-reconstruction-quality}, the more energy is
concentrated in the known part the worse is the reconstruction quality
of the unknown part. This phenomenon is, of course, expected. The
Poisson noise is signal dependent: higher intensity results in more
noise. However, the intensity (energy) of the unknown part remains
constant, hence the noise becomes more and more significant compared
to it. To obtain the best result one would like to design a mask whose
power spectrum will correlate well with the power spectrum of the
sought signal. Unfortunately, this approach cannot be implemented,
because designing such a mask requires a priori
knowledge of the sought signal's Fourier magnitude, which is
unavailable in our case. However, it may be a good approach when the
Fourier magnitude of the sought signal is known \emph{approximately}. 

\section{Concluding remarks}
\label{sec:boundary-conclusions}

In this chapter we considered the problem often met in the Fourier
domain holography: signal reconstruction form the Fourier magnitude of
the sum of the sought signal and a reference beam. We provided an
explanation to the fact observed in practice: why a strong reference
beam leads to a faster reconstruction for a variety of reconstruction
methods. Based on this explanation we suggested a ``good''  boundary
(reference beam) design. The latter problem (reference beam design)
requires more research, as the optimal reference beam must satisfy at
least two requirements in the presence of signal dependent noise. For
example, in the case of Poissonian noise (or any other noise model in
which the noise level grows with the signal intensity) the optimal
reference beam must be simultaneously ``strong'' (to aid the reconstruction process)
and ``weak'' (to alleviate) the destructive influence of the noise.

In general, when the Fourier magnitude of the sought image is known
approximately, the best mask should have the power spectrum that is
about two times larger than that of the sought signal (in each
frequency). If the power spectrum of the sought images is unknown, the
mask should have a strong presence in all frequencies. In this case,
it seems that the best design would be based on some ``randomness'' in
the mask values or shapes.

\bibliographystyle{osajnl}
\bibliography{My_Library}

\begin{thebibliography}{1}

\bibitem{bruck79ambiguity}
Yu.~M. Bruck and L.~G. Sodin.
\newblock On the ambiguity of the image reconstruction problem.
\newblock {\em Optics Communications}, 30(3):304--308, September 1979.

\bibitem{fiddy83enforcing}
M.~A. Fiddy, B.~J. Brames, and J.~C. Dainty.
\newblock Enforcing irreducibility for phase retrieval in two dimensions.
\newblock {\em Optics Letters}, 8(2):96--98, February 1983.

\bibitem{fienup82phase}
J.~R. Fienup.
\newblock Phase retrieval algorithms: a comparison.
\newblock {\em Applied Optics}, 21(15):2758--2769, 1982.

\bibitem{fienup83reconstruction}
J.~R. Fienup.
\newblock Reconstruction of objects having latent reference points.
\newblock {\em Journal of the Optical Society of America}, 73(11):1421--1426,
  November 1983.

\bibitem{fienup86phase}
J.~R. Fienup.
\newblock Phase retrieval using boundary conditions.
\newblock {\em Journal of the Optical Society of America A}, 3(2):284--288,
  February 1986.

\bibitem{hayes82reconstruction}
M.~Hayes.
\newblock The reconstruction of a multidimensional sequence from the phase or
  magnitude of its fourier transform.
\newblock {\em Acoustics, Speech, and Signal Processing [see also {IEEE}
  Transactions on Signal Processing], {IEEE} Transactions on}, 30(2):140--154,
  1982.

\bibitem{hayes82importance}
M.~Hayes and T.~Quatieri.
\newblock The importance of boundary conditions in the phase retrieval problem.
\newblock In {\em Acoustics, Speech, and Signal Processing, {IEEE}
  International Conference on {ICASSP} '82.}, volume~7, pages 1545--1548, 1982.

\bibitem{hayes83recursive}
Monson~H. Hayes and Thomas~F. Quatieri.
\newblock Recursive phase retrieval using boundary conditions.
\newblock {\em Journal of the Optical Society of America}, 73(11):1427--1433,
  November 1983.

\bibitem{osherovich11approximate}
Eliyahu Osherovich, Michael Zibulevsky, and Irad Yavneh.
\newblock Approximate {{Fourier}} phase information in the phase retrieval
  problem: what it gives and how to use it.
\newblock {\em Journal of the Optical Society of America A}, 28(10):2124--2131,
  October 2011.

\end{thebibliography}

\end{document}